\documentclass[twocol]{ametsoc}
\usepackage{ulem}

\journal{jas}

\bibpunct{(}{)}{;}{a}{}{,}

\title{Quantifying the eddy-jet feedback strength of the annular mode in an idealized GCM and reanalysis data}

    \authors{Ding Ma\correspondingauthor{Ding Ma, 
     20 Oxford St., Cambridge, MA 02138.} \thanks{Current Affiliation: Earth Institute, Columbia University, New York, NY}}
     \affiliation{\small Department of Earth and Planetary Sciences, Harvard University, Cambridge, Massachusetts}
\email{mading.pku@gmail.com}

    \extraauthor{Pedram Hassanzadeh}
    \extraaffil{\small Department of Mechanical Engineering, Rice University, Houston, Texas and Department of Earth and Planetary Sciences, Harvard University, Cambridge, Massachusetts}

    \extraauthor{Zhiming Kuang}
    \extraaffil{\small Department of Earth and Planetary Sciences, and School of Engineering and Applied Sciences, Harvard University, Cambridge, Massachusetts}

\abstract{
A linear response function (LRF) that relates the temporal tendency of zonal mean temperature and zonal wind to their anomalies and external forcing is used to accurately quantify the strength of the eddy-jet feedback associated with the annular mode in an idealized GCM. Following a simple feedback model, the results confirm the presence of a positive eddy-jet feedback in the annular mode dynamics, with a feedback strength of 0.137 day$^{-1}$ in the idealized GCM. 
Statistical methods proposed by earlier studies to quantify the feedback strength are evaluated against results from the LRF. It is argued that the mean-state-independent eddy forcing reduces the accuracy of these statistical methods because of the quasi-oscillatory nature of the eddy forcing. A new method is proposed to approximate the feedback strength as the regression coefficient of low-pass filtered eddy forcing onto low-pass filtered annular mode index, which converges to the value produced by the LRF when timescales longer than 200 days are used for the low-pass filtering. Applying the new low-pass filtering method to the reanalysis data, the feedback strength in the Southern annular mode is found to be 0.121 day$^{-1}$, which is presented as an improvement over previous estimates. This work also highlights the importance of using sub-daily data in the analysis by showing the significant contribution of medium-scale waves of periods less than 2 days to the annular mode dynamics, which was under-appreciated in most of previous research. The present study provides a framework to quantify the eddy-jet feedback strength in models and reanalysis data.
} 
\begin{document}
\maketitle
\section{Introduction}

The annular mode is a dominant mode of variability of the extratropical circulation in both hemispheres on intraseasonal to interannual timescales \citep{Kidson1988, Thompson1998, Gong1999, Thompson2000}. 
The annular mode corresponds to the leading empirical orthogonal function (EOF) of zonal mean zonal wind, which features an equivalent barotropic dipolar structure and represents latitudinal shifts of the eddy-driven midlatitude jet \citep{Nigam1990, Hartmann1998, Thompson2014,Thompson2015}. 
The zonal index, the time series associated with the annular mode, is essentially the same concept as that discussed in the pioneering studies of the variability of the general circulation \citep{Rossby1939, Namias1950, Wallace1985}. 
The annular mode in the Northern Hemisphere is often considered in recent studies as the hemispheric manifestation of the North Atlantic Oscillation \citep[e.g.,][]{Wallace2000, Vallis2004}. 
The annular mode is characterized by temporal persistence \citep{Baldwin2003, Gerber2008a, Gerber2008b}, for which it has been suggested that a positive feedback between anomalous zonal flow and eddy fluxes is responsible \citep[e.g.,][hereafter, LH01]{Feldstein1998, Robinson2000, Gerber2006, Lorenz2001}. 
For example, \citet{Robinson2000} suggested that at the latitudes of a positive anomaly of barotropic zonal wind, while surface drag tends to slow down low-level westerlies, it also enhances baroclinicity, which leads to stronger eddy generation. 
When the eddies propagate away, in the upper troposphere, from the latitudes where they are generated, the associated anomalies of eddy momentum flux reinforce the original zonal wind anomaly. 
As another example, \citet{Gerber2006} argued that anomalous baroclinicity is not necessarily required for a positive eddy-jet feedback, as the mean flow anomaly can change the position of the critical latitudes for wave breaking and influence the eddy momentum flux convergence.

Quantifying the strength of eddy-jet feedback is important for understanding both internal variability and response to external forcing. 
One common issue with the current GCMs is that the simulated annular mode is too persistent compared to observations \citep{Gerber2008a}, which not only indicates biases of jet variability, but also suggests overestimation of changes in the extratropical circulation in response to anthropogenic forcing in the models. 
According to the fluctuation-dissipation theorem \citep{Leith1975}, the magnitude of the forced response is positively related to the timescale of the unforced variability, a relationship that has been confirmed qualitatively in some atmospheric models \citep[e.g.,][]{Ring2008, Chen2009}.

Based on the assumption that the mean-state-independent eddy forcing does not have long-term memory, LH01 and \citet[][hereafter, S13]{Simpson2013b} attributed positive values of lagged correlations between the zonal index and the eddy forcing, when the zonal index leads eddy forcing by a few days, to a positive feedback, and proposed statistical methods to quantify the strength of eddy-jet feedback in observations and simulations to improve understanding of the persistence of the jet. {Even though} S13 validated their method using synthetic time series generated by a second-order autoregressive process, {their} statistical method{, as well as the statistical method proposed by LH01, would benefit from an assessment} with more realistic time series of zonal index and eddy forcing. 
Due to the chaotic nature of eddies, the mean-state-dependent eddy forcing cannot be separated from the mean-state-independent part in the reanalysis data, and as a result, it is difficult to validate the assumptions of these statistical methods. Furthermore, a recent study showed that the existence of an internal eddy feedback cannot be distinguished from the presence of external interannual forcing using only the statistical methods (Byrne et al. 2016).

In the present study a linear response function (LRF), following \citet{Pedram2016a}, is used to identify the anomalous eddy fluxes in response to mean state anomalies that match the spatial pattern of annular mode in an idealized GCM. 
This provides the ``ground truth" in the idealized GCM, and serves as a benchmark against which one can assess the statistical methods. 
The LRF will be briefly explained in Section 2, along with model configuration and the reanalysis data. 
In Section 3, the annular mode and a simple model of eddy-jet feedback will be introduced, followed by quantification of the feedback strength using different methods in Section 4. 
Discussions and a brief summary are presented in Section 5.

\section{Methodology}

For the numerical simulations, we use the Geophysical Fluid Dynamics Laboratory dry dynamical core, which solves the primitive equations with Held-Suarez forcing \citep{Held1994}. 
Temperature is relaxed to an equinoctial radiative-equilibrium state with an equator-to-pole temperature difference of 60 K. Similar setups have been widely used to study the midlatitude circulation and its low-frequency variability \citep[e.g.,][]{Gerber2008b, Chen2009, Hassanzadeh2014, Hassanzadeh2015, McGraw2016}.
Each simulation is integrated for 45000 days at the T63 resolution (horizontal spacing of around 200 km) with 40 vertical levels and 6-hourly outputs, and the first 500 days are discarded. 
Ten ensemble simulations are conducted for the control (CTL) and an experiment (EXP). 
In EXP, a zonally symmetric time-invariant forcing is applied to zonal wind and temperature, so that the difference of the equilibrium mean states between EXP and CTL matches the pattern of the annular mode in CTL. 
This external forcing is calculated using the LRF found by \citet{Pedram2016a}, and EXP is essentially the same as Test 3 in their article. The LRF ($\mathbf{L}$ in Equation 1) relates anomalous state vector $\mathbf{x}$ to its temporal tendency and an external forcing $\mathbf{f}$ as, 

\begin{equation}
\renewcommand{\theequation}{1}
\frac{d\mathbf{x}}{dt} = \mathbf{L} \mathbf{x} + \mathbf{f},
\end{equation}

\noindent in which $\mathbf{x}$ consists of $[\mathbf{u}]$ and $[\mathbf{T}]$, zonally averaged (denoted by square brackets) zonal wind and temperature anomalies from the mean state of CTL. 
Assuming that eddies are in statistical equilibrium with the mean flow in the long-term integrations, Equation 1 is valid for weak external forcings (see Hassanzadeh and Kuang 2016a for more details). 
With $\mathbf{x}_{o}$ denoting the anomalous state vector associated with the annular mode, the particular external forcing for EXP is $\mathbf{f}_{o} = -\mathbf{L} \mathbf{x}_{o}$.

{It is worth mentioning that Hassanzadeh and Kuang (2016a) have shown that the leading EOF of [u] and [T] strongly resembles the singular vector of the LRF that has the smallest singular number} \citep[the so-called neutral vector, see][]{Goodman2002}, {which confirms that the annular mode is indeed a dynamical mode, rather than a statistical artifact, in the idealized GCM. 
They further argued that given the similarities between the annular mode in the real atmosphere and the one simulated in the idealized GCM, it is plausible that the annular mode is also the neutral vector and hence a real dynamical mode of the real atmosphere (and atmospheres modeled with more complex GCMs), which can explain the ubiquity of annular-mode-like responses in the forced atmospheric circulations.}

For the observational analysis, National Centers for Environmental Prediction reanalysis 2.5$^\circ$ latitude $\times$ 2.5$^\circ$ longitude 6-hourly wind and temperature from 1951 to 2014 are used. 
Anomalies are calculated by removing the annual average and the first four Fourier harmonics as in LH01. 
Following \citet{Baldwin2009}, spatial weighting is applied to EOF analysis and projections of spatial patterns to compensate for the uneven distribution of grids in both model outputs and reanalysis data. 
For spectral analyses, input data is divided into 1024-day segments unless otherwise noted.

{Here, we emphasize that 6-hourly data, rather than daily data, is used in the present study in order to capture the medium-scale waves} \citep{Sato2000}. 
{It has been shown that the medium-scale waves, which have timescales shorter than 2 days, play an important role in the annular mode dynamics despite their weak climatological amplitudes} \citep{Kuroda2011}. 

\section{Annular mode and eddy-jet feedback}

\subsection{Jet climatology and annular mode structure}
{We} will be focusing on the Southern annular mode in the reanalysis data {for simplicity, considering the longitudinal symmetry in the Southern Hemisphere}. There are two separate jets in the Southern Hemisphere {climatology} (Figure 1a), namely, the subtropical jet centering around 35$^\circ$S and the midlatitude jet at around 50$^\circ$S.
{Strictly following LH01, the zonal index is defined as the leading principal component (PC) of $\langle [\mathbf{u}]\rangle$, in which the angle brackets denote vertical average. 
The leading EOF of $\langle [\mathbf{u}]\rangle$ explains 40$\%$ of the total variance, while the second EOF explains 20$\%$. Here the zonal index is normalized so that its standard deviation is one. 
The latitude-pressure pattern of $[\mathbf{u}]$ and $[\mathbf{T}]$ associated with the annular mode in the reanalysis data can be seen by regressing $[\mathbf{u}]$ and $[\mathbf{T}]$ on the zonal index at zero-day lag (Figures 1bc). Note that the correlation between the zonal index and the leading PC of $[\mathbf{u}]$ is 0.995, so Figure 1b is essentially equivalent to the leading EOF of $ [\mathbf{u}]$.} 
The anomalous zonal mean zonal wind associated with the annular mode is characterized by an equivalent barotropic dipole, which is, as expected, in thermal wind balance with the zonal mean temperature anomaly. Variations in the zonal index represent {north-south} vacillations of the eddy-driven jet \citep[e.g.,][]{Hartmann1998}. 

\begin{figure*}[t]
 \centerline{\includegraphics[width=\textwidth]{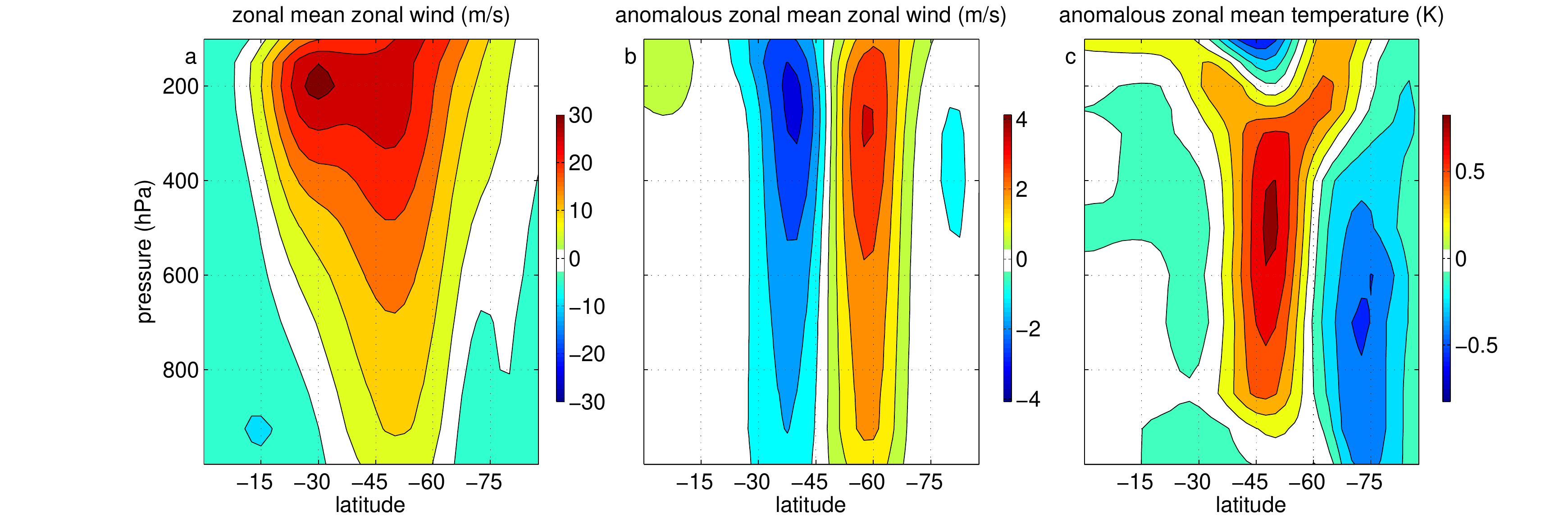}}
  \caption{(a) Climatology of zonal mean zonal wind in the reanalysis data. Anomalous (b) zonal mean zonal wind and (c) zonal mean temperature regressed on the leading PC of $\langle [u] \rangle$.}
  \label{fig1}
\end{figure*}

For model outputs, both hemispheres are analyzed, but the Northern Hemisphere is flipped and plotted as the Southern Hemisphere, as the model is symmetric about the equator. 
The climatology in the simulations with the same model configuration has been well documented \citep[e.g.,][]{Held1994}. 
In brief, a confined midlatitude jet centering around 40$^\circ$S, 10$^\circ$ equatorward to the eddy-driven jet in the reanalysis data, is produced in the CTL (Figure 2a). 
The zonal index is again calculated as the leading PC of $\langle [\mathbf{u}]\rangle$. 
The leading EOF of $\langle [\mathbf{u}]\rangle$ explains 50$\%$ of the total variance in the model, while the second EOF explains 18$\%$. Despite the idealized nature of the GCM, {the tropospheric dipolar} pattern of zonal wind of the annular mode produced in the model compares reasonably well with the Southern annular mode in the reanalysis data (Figures 2bc). 

\begin{figure*}[t]
 \centerline{\includegraphics[width=\textwidth]{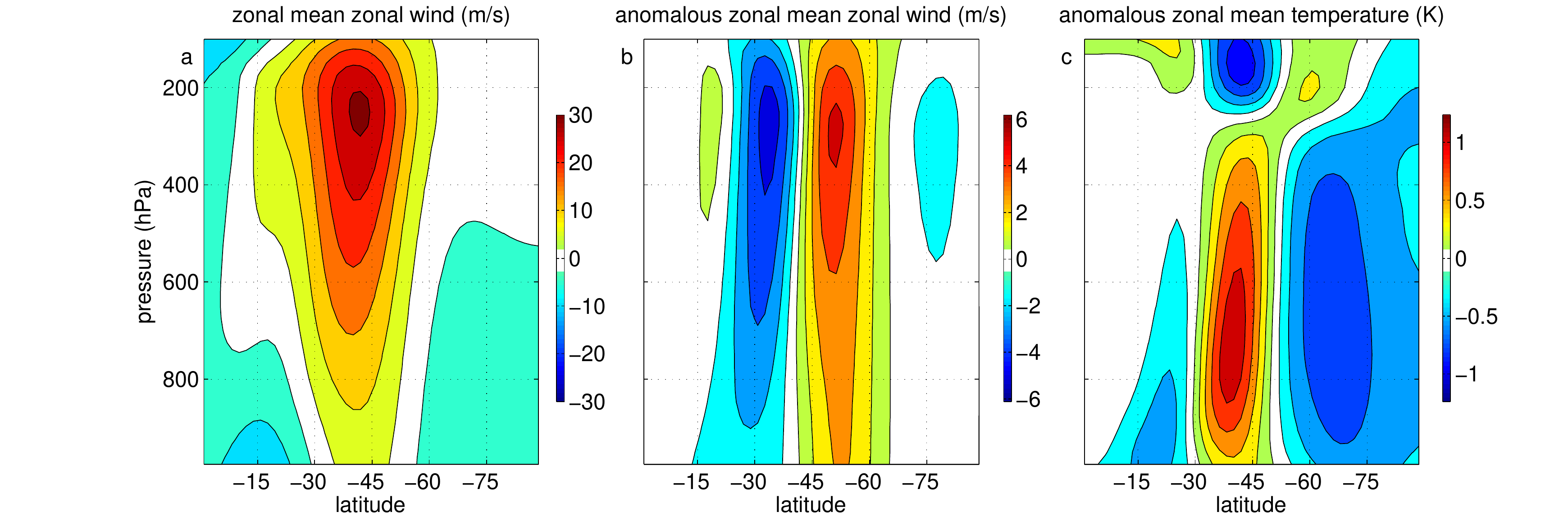}}
  \caption{The same as Figure 1, except for model outputs of CTL.}
  \label{fig2}
\end{figure*}

\subsection{Simple model of feedback}
{In their seminal work}, LH01 introduced a simple model of the eddy-jet feedback, which will be briefly explained in this section. With the same notations as in LH01, $z(t)$ indicates the zonal index, and $m(t)$ denotes the time series of eddy forcing on the annular mode, which is defined as the projection of the anomalous eddy momentum convergence onto the leading EOF of $\langle [\mathbf{u}]\rangle$. 
As discussed in LH01, the tendency of $z$ is formulated as,
\begin{equation}
\renewcommand{\theequation}{2a}
 dz/dt = m - z/\tau,
 \end{equation}
 
 \noindent in which $\tau$ is the damping timescale. Equation 2a can be interpreted as the zonally and vertically averaged zonal momentum equation (LH01),
\begin{equation*}
 \frac{\partial \langle[u]\rangle}{\partial t} = \frac{1}{\cos^2 \phi} \frac{\partial(\langle[u'v']\rangle \cos^2 \phi)}{a \partial \phi} - F,
 \end{equation*}
\noindent {where $u'$ and $v'$ are deviations of zonal wind and meridional wind from their respective zonal means, $\phi$ is the latitude, $a$ is the Earth's radius, and $F$ includes the effects of surface drag and secondary circulation.}

With capital letters denoting the Fourier transform of the corresponding lower case variables and $\omega$ denoting {angular} frequency, Equation 2a can be written as, 

\begin{equation}
\renewcommand{\theequation}{2b}
i\omega Z = M - {Z}/{\tau}
\end{equation}

Figure 3a shows the power spectrum of the zonal index in the reanalysis data, with a lowest resolved frequency of 1/1024 cycles per day (cpd). 
The zonal index features increasing power with decreasing frequency. 
At intraseasonal and shorter timescales, where the dominant balance of Equation 2b is between $i\omega Z$ and $M$, the power spectrum of zonal index can be interpreted, to the first order, as reddening of the power spectrum of eddy forcing (Figure 3b). 
The broad peak at synoptic timescale in the power spectrum of eddy forcing (Figure 3c) is an intrinsic characteristic of the mean-state-independent eddies (LH01).
At timescales longer than around 50 days, a positive eddy-jet feedback is suggested to be responsible for the high power of both of the zonal index and eddy forcing, where the dominant balance of Equation 2b is between $Z/\tau$ and $M$. 
{A linear feedback model for} $M$ (e.g., Hasselmann 1976; LH01) can be {written as},

\begin{equation}
\renewcommand{\theequation}{3}
M = \tilde{M} + bZ,
\end{equation}

\noindent where $\tilde{M}$ is the mean-state-independent eddy forcing, and $b$ is the strength of the eddy-jet feedback. 
In equilibrium, $b$ must be smaller than $1/\tau$ in both GCMs and the realistic atmosphere, otherwise the zonal index grows unboundedly. 
Plugging Equation 3 into Equation 2b returns,

\begin{equation}
\renewcommand{\theequation}{4}
i\omega Z = \tilde{M}  + (b - 1/{\tau})Z
\end{equation}

\noindent If we consider $\tilde{M}$ as {white noise at low frequencies}, the amplitude of $Z$ is inversely proportional to the difference between $1/\tau$ and $b$ at the low-frequency limit (i.e., neglecting the left hand side of Equation 4). 
The stronger the eddy feedback is (i.e., the closer $b$ is to $1/\tau$), the higher power $Z$ has at intraseasonal and longer timescales. 
Note that if $b=0$, the amplitude of $Z$ is inversely proportional to $1/\tau$ at the low-frequency limit, and at intraseasonal to interannual timescales the zonal index will still have increasing power with decreasing frequency \citep{Hasselmann1976}, although the annular mode will be less persistent than that with a positive eddy feedback.

\begin{figure*}[t]
 \centerline{\includegraphics[width=\textwidth]{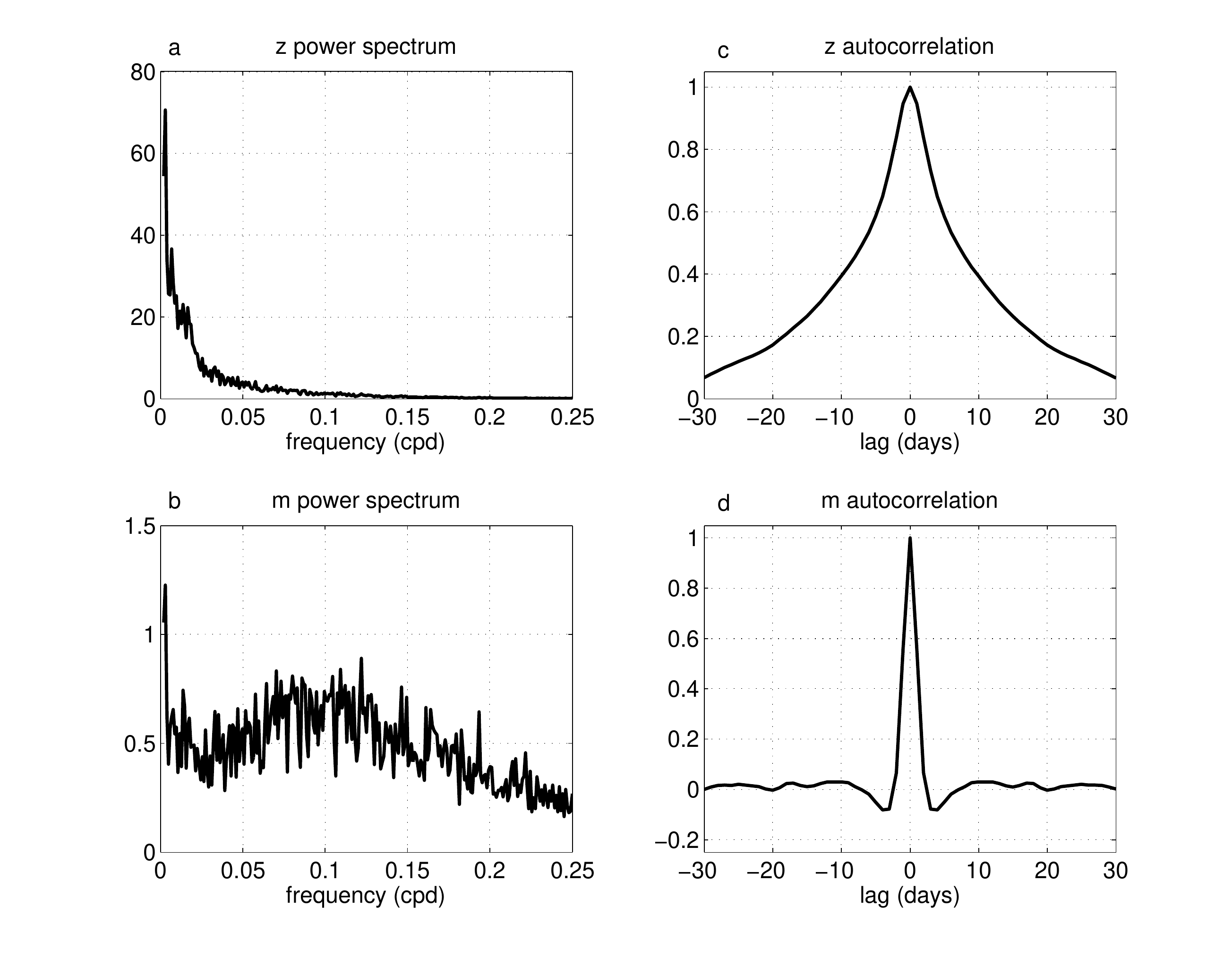}}
  \caption{Summary statistics for $z$ and $m$ in the reanalysis data. Power spectrum of (a) $z$ and (b) $m$, and autocorrelations of (c) $z$ and (d) $m$.}
  \label{fig3}
\end{figure*}

The autocorrelation function of the zonal index decreases more slowly with lag time than that of the eddy forcing (Figure 3cd). 
The negative autocorrelations of eddy forcing at small lag time indicates the quasi-oscillatory nature of the eddies (Figure 3d), which is consistent with the broad maximum in the power spectrum at 7-15 days. 
The cross-correlation of $m$ and $z$ peaks at around 0.53, when the zonal index lags eddy forcing by 1-2 days as the zonal index is driven by the eddy forcing (Figure 4). 
Negative cross-correlations when the zonal index leads eddy forcing by a fews days result from the oscillatory behavior of eddy forcing, and positive values at large lags suggest a positive eddy-jet feedback according to LH01. 

\begin{figure*}[t]
 \centerline{\includegraphics[width=0.5\textwidth]{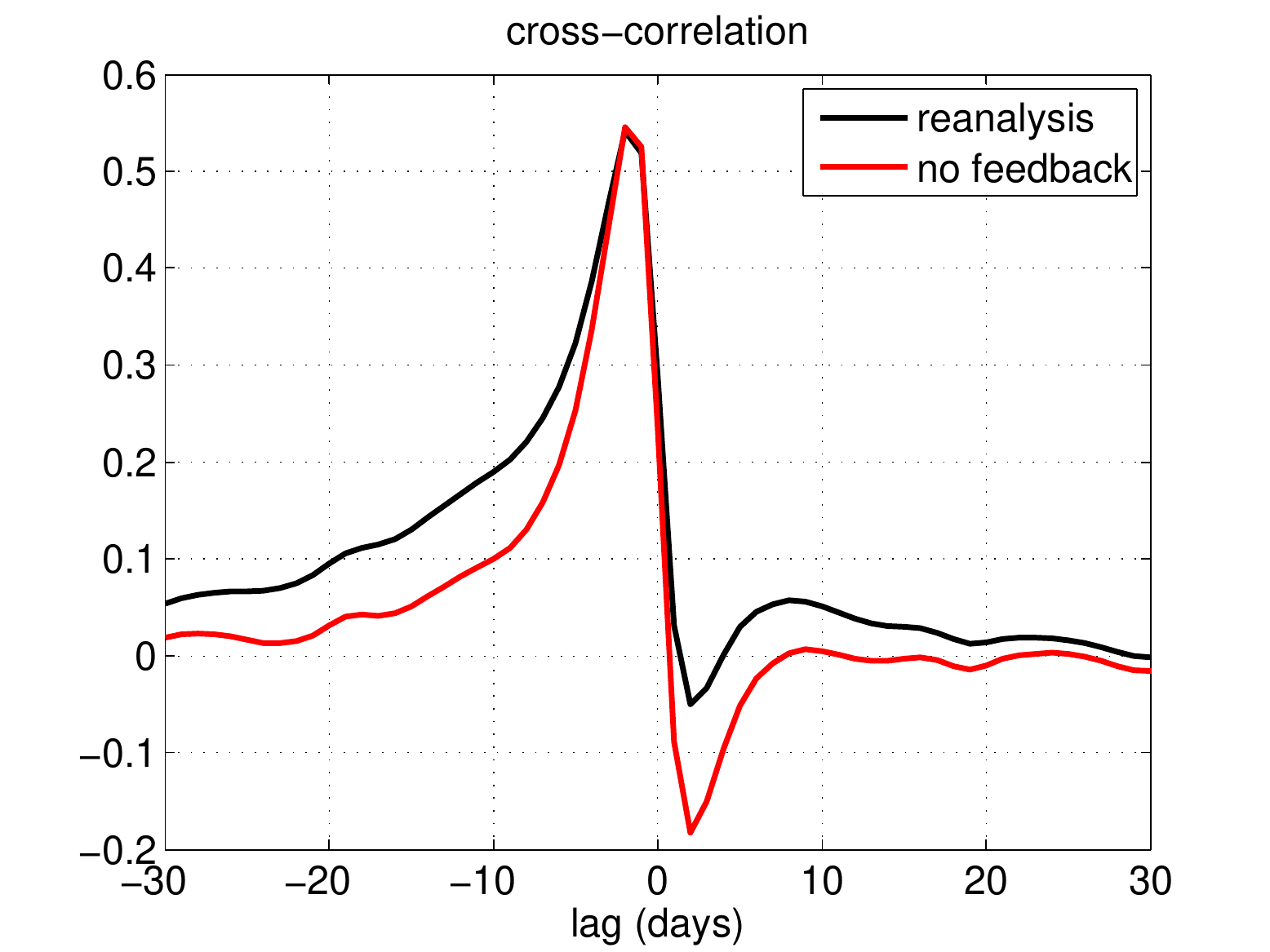}}
  \caption{Cross-correlation between $z$ and $m$ in the reanalysis data (black curve), and between $\tilde{z}$ and $\tilde{m}$ (i.e., without eddy feedback following LH01). Positive values of lag denote that zonal index leads eddy forcing.}
  \label{fig4}
\end{figure*}

Despite some biases, the CTL is able to capture the general features of the system as in the reanalysis data described above (Figure 5). 
The broad peak of eddy forcing at synoptic timescales in the power spectrum is more pronounced in the model, which indicates that the eddy forcing is more oscillatory in the idealized GCM. 
\citet{Chen2009} argued that the shoulders in the autocorrelation function of the zonal index at around $\pm$4-day lag can be attributed to the strong oscillatory nature of eddy forcing in the idealized GCM.
Also, the annular mode is more persistent in this GCM, as the cross-correlation between $m$ and $z$ decays more slowly compared to that in the reanalysis data (Figures 4 and 6), or equivalently, the simulated zonal index has higher power at intraseasonal and longer timescales compared to that in the reanalysis data.
Note that this is not just a bias of this idealized GCM. 
Too persistent annular modes are seen in GCMs of varying degrees of complexity, the cause of which is unknown and remains an important topic of research \citep{Gerber2008a, Gerber2008b, Nie2014}.

\begin{figure*}[t]
 \centerline{\includegraphics[width=\textwidth]{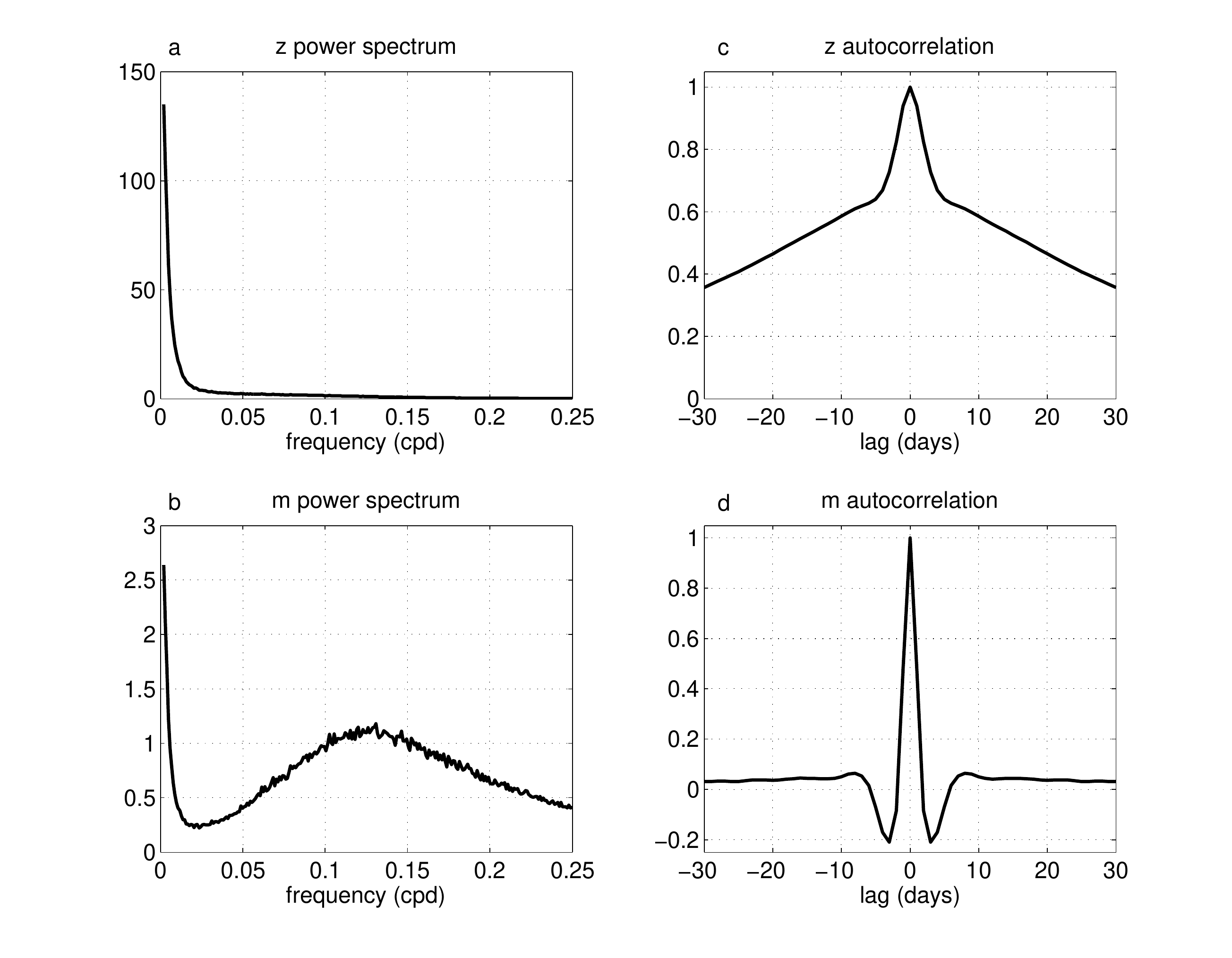}}
  \caption{The same as Figure 3, except for model outputs of CTL.}
  \label{fig5}
\end{figure*}

\begin{figure*}[t]
 \centerline{\includegraphics[width=0.5\textwidth]{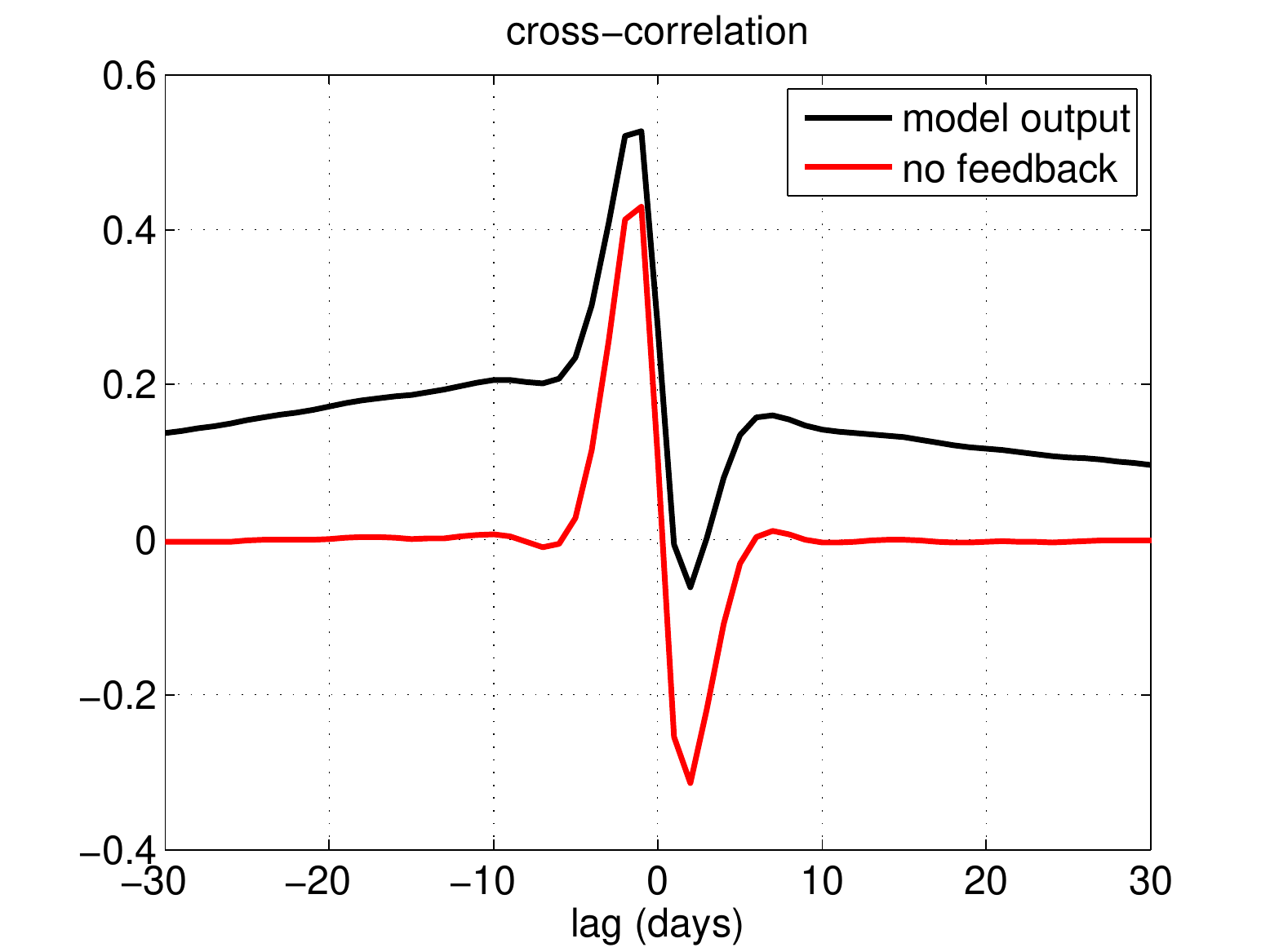}}
  \caption{The same as Figure 4, except for model outputs of CTL.}
  \label{fig6}
\end{figure*}

\section{Eddy-jet feedback strength}

The LRF will first be used to calculate the ``ground truth" of the eddy-jet feedback strength associated with the leading EOF of $\langle [u] \rangle$ (i.e., the annular mode), as well as the second EOF, in the idealized GCM. 
Three different statistical methods, namely, fitting cross-correlation functions (LH01), lag regression (S13) and regression using low-pass filtered data (introduced in the present study), will be used to estimate the eddy feedback strength of the annular mode in the idealized GCM, and evaluated against the result from the LRF. 
Then we will apply the statistical methods to investigate the eddy feedback associated with the annular mode in the reanalysis data.

\subsection{Linear response function}
With a zonally symmetric time-invariant forcing, the deviations of mean state in EXP from that in CTL (Figures 7ab) are nearly identical to the pattern of the annular mode (Figures 2bc){, with a pattern correlation of 0.995}. 
Note that the changes in the mean state from CTL to EXP are caused by the imposed external forcing and are long term averages so that the eddies are in statistical equilibrium with the mean state. 
The changes of eddy fluxes from CTL to EXP are the response to the mean state changes, rather than the cause of the deviation of the mean state.
The anomalous eddy fluxes are shown in Figures 7cd, the pattern of which largely agrees with LH01. 
In the region of positive zonal wind anomalies (around 50$^\circ$), meridional temperature gradient increases at low levels (Figures 7ab), leading to enhanced baroclinic wave generation and stronger eddy heat flux (Figure 7d). 
Correspondingly, the equatorward propagation of waves enhances the poleward eddy momentum flux at around 45$^\circ$, which reinforces the zonal wind anomaly (Figure 7c). 
The strength of the eddy feedback can be calculated by projecting the anomalous eddy momentum flux convergence onto the anomalous zonal wind (see Baldwin et al. 2009 for details about projection of data with spatial weighting).
The averaged feedback strength of the 10 ensemble simulations (referred to as $b_{LRF}$ hereafter) is around 0.137 day$^{-1}$, which is denoted by the red solid line in Figure 8. 
The red dashed lines in Figure 8 show the 95$\%$ confidence intervals of $b_{LRF}$, indicating little spread across the ensemble members.
$b_{LRF}$ is considered as the ground truth in the idealized GCM.

\begin{figure*}[t]
 \centerline{\includegraphics[width=\textwidth]{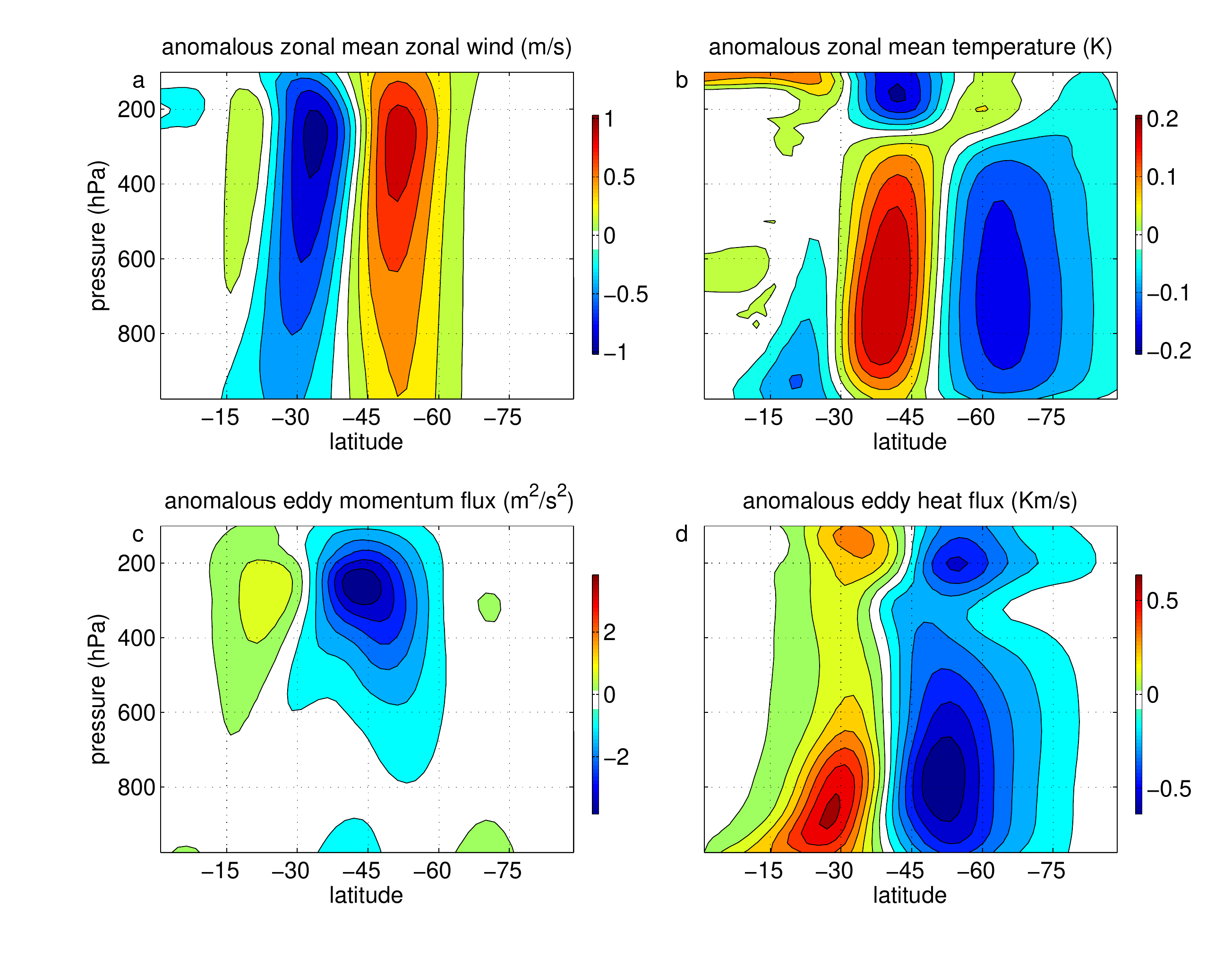}}
  \caption{The difference of (a) zonal mean zonal wind, (b) zonal mean temperature, (c) zonal average eddy momentum flux and (d) zonal average eddy heat flux between EXP and CTL.}
  \label{fig7}
\end{figure*}

\begin{figure*}[t]
 \centerline{\includegraphics[width=\textwidth]{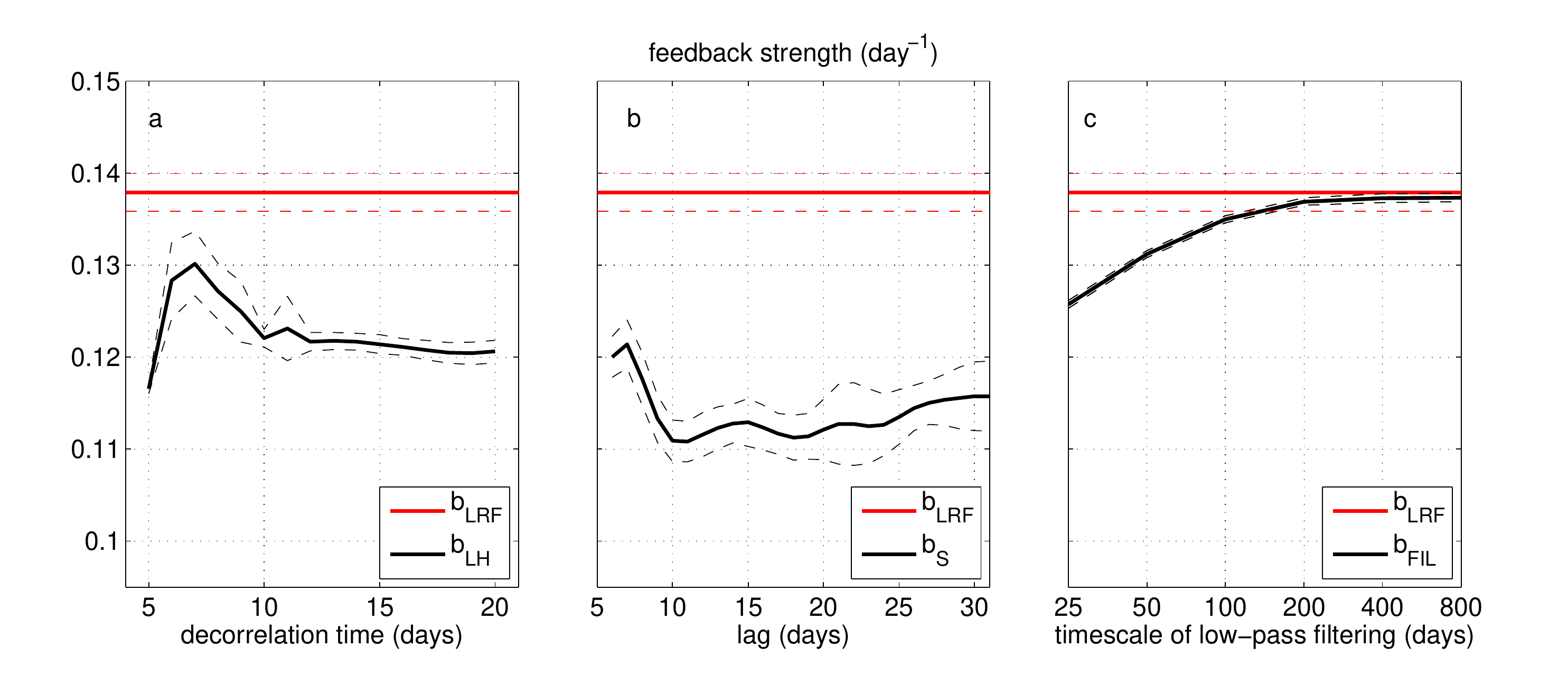}}
  \caption{Strength of eddy-jet feedback estimated in the idealized GCM following different methods: (a) LH01, (b) S13 and (c) low-pass filtering. The red lines in each panel shows the value calculated using the LRF. The dashed lines denoting 95$\%$ confidence intervals}
  \label{fig8}
\end{figure*}

The mean-state-independent eddy forcing is not directly observable and cannot be separated from the mean-state-dependent eddy forcing in the reanalysis data, but can be computed in the idealized GCM as $\tilde{M} = M - b_{LRF}Z$. 
The power spectrum of the mean-state-independent eddy forcing is shown in Figure 9. 
At timescales shorter than around 50 days, the mean-state-independent eddy forcing dominates the total eddy forcing. 
In particular, it is confirmed that the mean-state-independent eddy forcing is responsible for the broad peak of total eddy forcing at synoptic timescales. 
At timescales longer than 50 days, the strength of the mean-state-independent eddy forcing decreases with decreasing frequency, while the strength of the total eddy forcing rises as frequency decreases. 

\begin{figure*}[t]
 \centerline{\includegraphics[width=22pc]{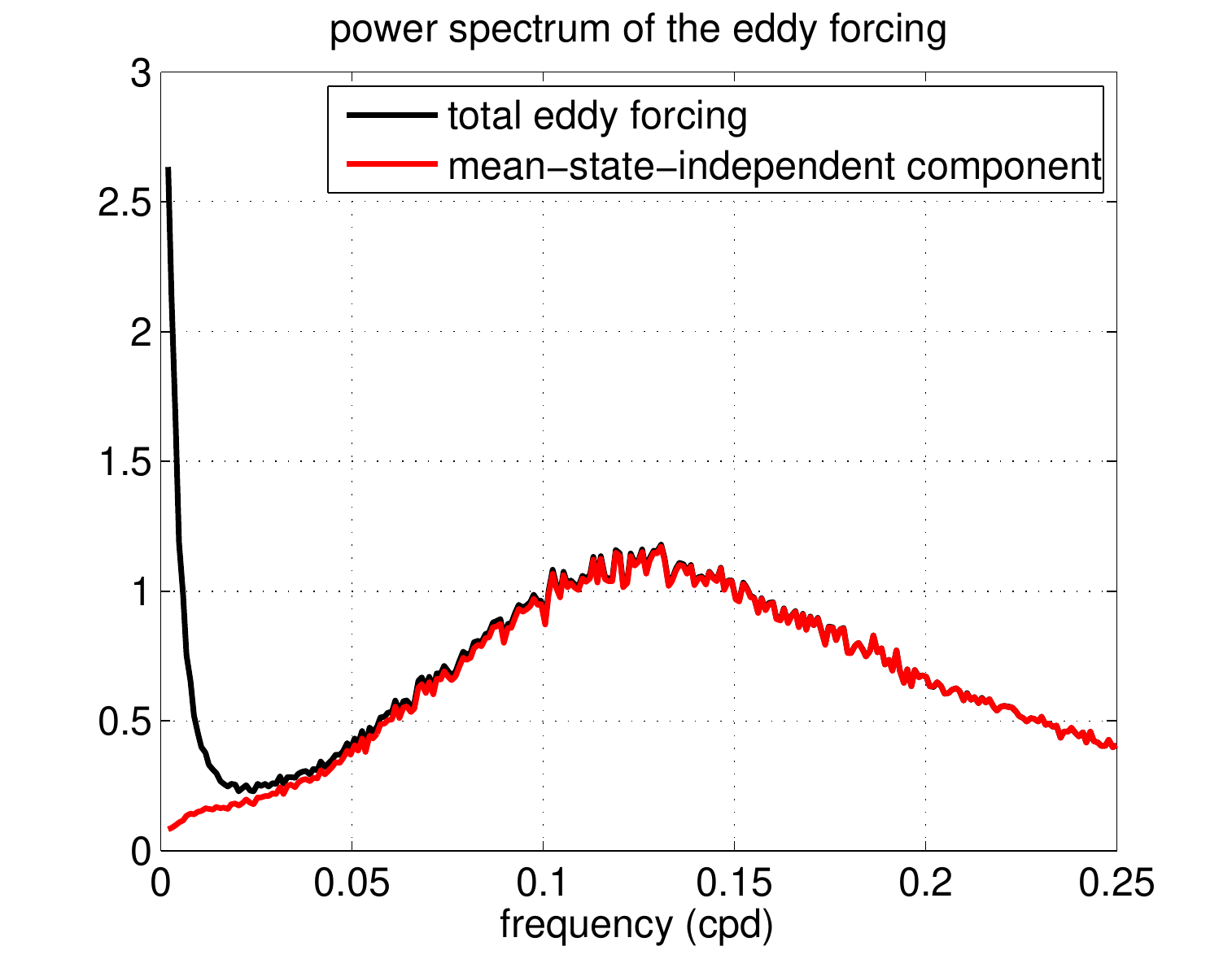}}
  \caption{Power spectrum of the total eddy forcing (black) and the mean-state-independent eddy forcing (red).}
  \label{fig9}
\end{figure*}

{At intraseasonal to interannual timescales, the total eddy forcing is dominated by mean-state-dependent eddy forcing. 
Here, the role of the medium-scale waves, whose period is shorter than 2 days, in the annular mode dynamics is emphasized. 
It has been shown that the amplitude of the medium-scale waves, which is weak in the climatology, is strongly modified by the annular mode, and the fluxes resulting from these waves have a substantial contribution to the annular mode dynamics} \citep{Kuroda2011}. 
At interannual timescales, the total eddy forcing calculated from daily wind anomalies captures less than half of the total eddy forcing calculated from 6-hourly wind anomalies in the idealized GCM (Figure 10a). 
The results suggest that the eddy-jet feedback will be strongly underestimated without accounting for medium-scale waves. 
In fact, with daily model outputs, $b_{LRF}$ is around 0.083 day$^{-1}$, 40$\%$ weaker than that calculated using 6-hourly model outputs.

\begin{figure*}[t]
 \centerline{\includegraphics[width=\textwidth]{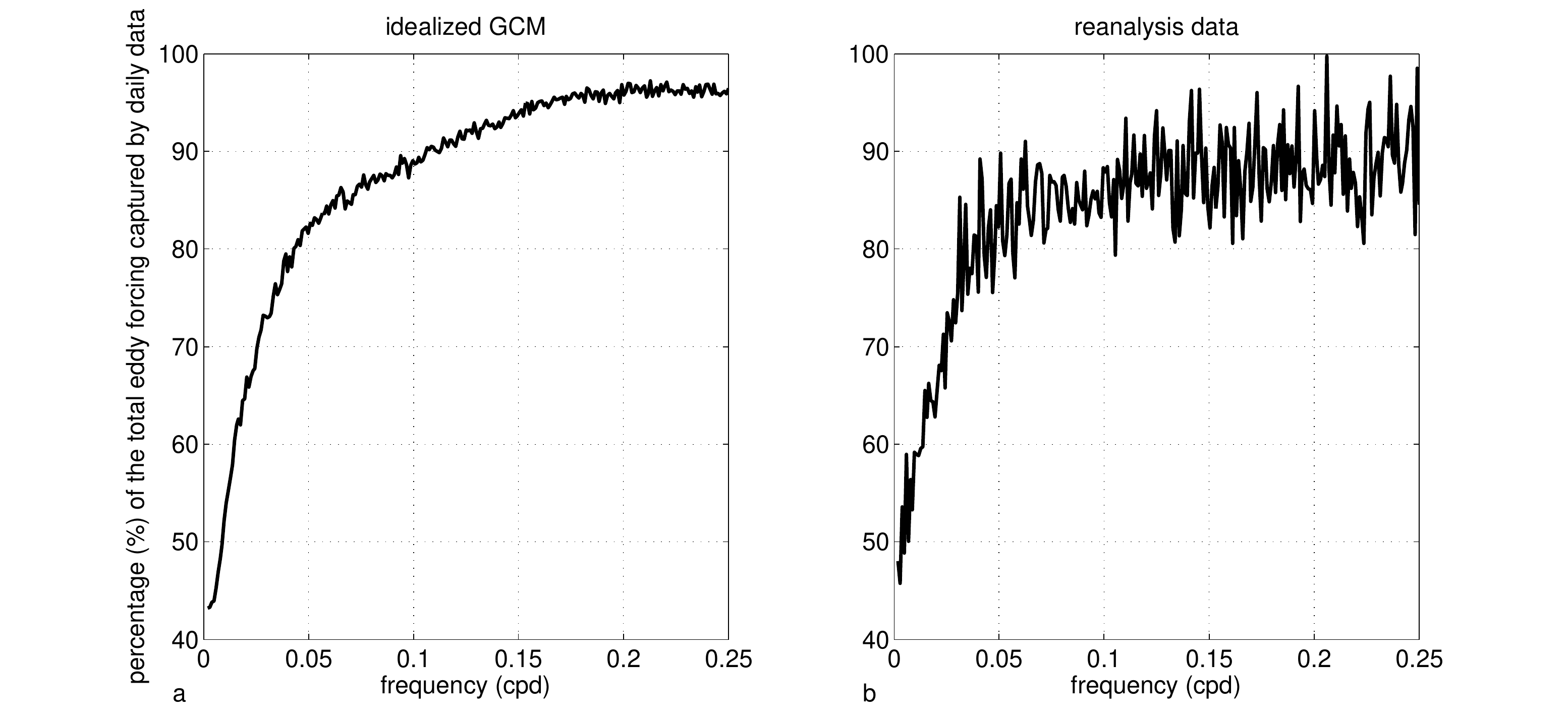}}
  \caption{The ratio between the total eddy forcing calculated from daily wind anomalies and that calculated from 6-hourly wind anomalies for (a) model outputs of CTL and (b) the reanalysis data.}
  \label{fig10}
\end{figure*}

{Although the focus of the present work is on the annular mode (i.e., the leading EOF of the zonal mean zonal wind), we also apply the LRF framework to the second EOF, which is characterized by a tripolar pattern of zonal wind anomalies and corresponds to the fluctuations of the amplitude of the jet (Figure 11a). 
With a stronger midlatitude jet, temperature gradient is enhanced between 30$^\circ$S-40$^\circ$S below around 300 hPa (Figure 11b). 
Poleward eddy heat flux is strengthened due to sharper temperature gradient (Figure 11d), and the anomalous eddy momentum flux associated with second EOF tends to export momentum out of the jet (Figure 11c).
Using another ensemble of 10 simulations with an external forcing calculated for the second EOF, it is found that the eddy feedback associated with the second EOF is negative, and the strength of the feedback is -0.264 day$^{-1}$. 
This is consistent with the findings of LH01, who inferred from a lag-regression analysis that the feedback is negative. 
LH01 also argued that the anomalous eddy momentum flux associated with the second EOF tend to weaken the jet as a result of increased barotropic shear, i.e. the barotropic governor effect} \citep{James1987}.

\begin{figure*}[t]
 \centerline{\includegraphics[width=\textwidth]{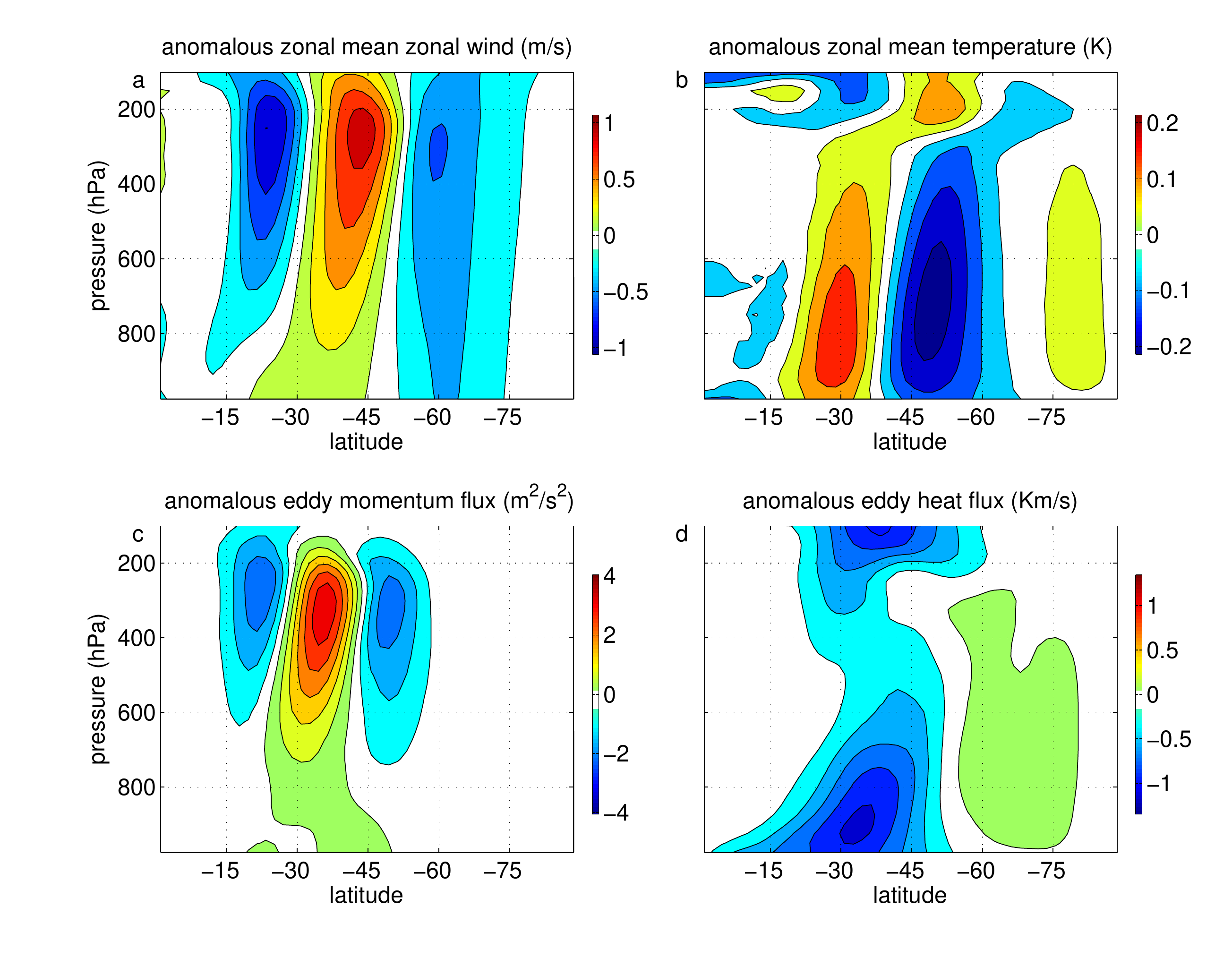}}
  \caption{The same as Figure 7, except for the second EOF of zonal mean zonal wind.}
  \label{fig11}
\end{figure*}

\subsection{Fitting cross-correlation functions (LH01)}
In a pioneering study, LH01 inferred the existence of a positive eddy-jet feedback in the annular mode dynamics from the reanalysis data and based on the the assumption that the mean-state-independent eddy forcing has short memory (i.e., the time series of $\tilde{m}$ has a short decorrelation timescale), and proposed the following method to quantify the strength of the feedback by fitting the covariance functions. If $b$ = 0, Equation 4 becomes,

\begin{equation}
\renewcommand{\theequation}{5}
i\omega \tilde{Z} = \tilde{M} - \tilde{Z}/{\tau},
\end{equation}

\noindent where $\tilde{Z}$ denotes the zonal index in a system without eddy-jet feedback. 
The covariance between $\tilde{z}$ and $\tilde{m}$ must be close to zero when $\tilde{z}$ leads $\tilde{m}$ by a period longer than the decorrelation timescale of the mean-state-independent eddies. 
It has been shown that the covariance between $\tilde{z}$ and $\tilde{m}$ is a function of $b$ and the covariance between ${z}$ and ${m}$ (see LH01 for details), and $b$ can be estimated by minimizing the mean squared cross-correlations at lags longer than a particular decorrelation timescale. 
For instance, assuming a decorrelation time of 7 days, the estimated strength of eddy-jet feedback (hereafter $b_{LH}$) is around 0.13 day$^{-1}$, and the red curve in Figure 6 shows the corresponding cross-correlations between $\tilde{z}$ and $\tilde{m}$. 
Bootstrap confidence intervals (at 95$\%$ confidence levels) are plotted to indicate errors (black dashed curves in Figure 8a). 
A bootstrap ensemble of 5000 members is constructed by resampling from the original time series. 
Feedback strength is calculated for each of the bootstrap ensemble member, which provides the probability density function of $b_{LH}$ and thus the confidence intervals. 
$b_{LH}$ varies with the choices of decorrelation time. Note that it is difficult to determine an optimal decorrelation time {\textit{a priori} due to the quasi-oscillatory behavior of $\tilde{m}$}, especially when the decorrelation timescale varies by season \citep[e.g.,][]{Sheshadri2016}.

\subsection{Lag regressions}
Lag regression is applied to find the feedback strength following S13. 
Denote the auto-covariance function of $z$ with lag $l$ as $\gamma_z (l)$, and write the cross-covariance function between $z$ and $m$ as $\gamma_{zm} (l)$ when $z$ leads $m$ by $l$ days.
Consider the lag regression model $m(t) = \beta(l)z(t-l) $, the lag regression coefficient $\beta$ is,
\begin{equation}
\renewcommand{\theequation}{6}
\beta(l) = \frac{\gamma_{zm}(l)}{\gamma_{z}(0)}
\end{equation}

With Equation 3, the right hand side of Equation 6 can be decomposed into two parts:
\begin{equation}
\renewcommand{\theequation}{7}
\beta(l) = \frac{\gamma_{z\tilde{m}}(l)}{\gamma_{z}(0)} +b\frac{\gamma_{z}(l)}{\gamma_{z}(0)},
\end{equation}
\noindent in which the first term on the right hand side {is negligible} if $z$ is decorrelated with $\tilde{m}$ beyond lag $l$ days, and therefore the feedback strength can be estimated as,

\begin{equation}
\renewcommand{\theequation}{8}
b_{S} = \beta(l) \frac{\gamma_{z}(0)} {\gamma_{z}(l)}
\end{equation}

Figure 8b shows the strength of eddy-jet feedback calculated using Equation 8, with 95$\%$ confidence intervals estimated with bootstrapping as in Section 4b. 
While the margin of error grows with lag time, the strength of eddy-jet feedback is largely underestimated, and the bias results from the quasi-oscillatory nature of the eddy forcing. 
Using lag regression, we are also able to estimate the pattern of anomalous eddy fluxes associated with the annular mode. 
The pressure-latitude distribution of eddy flux anomaly generally agrees with the results from LRF, with a pattern correlation over 0.9 through a wide range of lag days (figures not shown). 

\subsection{Low-pass filtering}

The bias with lag regression suggests that the correlation between $\tilde{m}$ and $z$ is not negligible {relative to the correlation between $m$ and $z$} at a lag as long as 30 days (Figure 8b). 
One can expect that at longer lag timescales, $\tilde{m}$ and $z$ eventually become decorrelated and thus Equation 8 will be valid, but it can also be expected that with such long lag time, the margin of error will be large so that the estimation is uninformative. 
Inspired by the observation that the strength of the mean-state-independent eddy forcing vanishes at the low-frequency limit (Figure 9), here we propose a new method to bypass this issue. Multiplied by ${Z^{*}}/{(ZZ^{*})}$ on both sides, where $Z^{*}$ denotes the conjugate of $Z$, Equation 3 becomes:

\begin{equation}
\renewcommand{\theequation}{9}
\frac{MZ^*}{ZZ^*} = \frac{\tilde{M}Z^*}{ZZ^*} + b
\end{equation}

Using the LRF, the real component of the first term on the right hand side can be explicitly calculated and is found to be negligible at the low-frequency limit (Figure 12). 
As a result, the feedback strength equals the real component of the left hand side of Equation 9 at the lowest frequencies, which can be calculated as the regression coefficient of low-pass filtered $m$ on low-pass filtered $z$. 
{In practice, Lanczos filtering is applied with the number of weights covering the length of four times of the cut-off periods.} 
The estimated feedback strength (denoted as $b_{FIL}$) is plotted in Figure 8c. When timescales longer than 200 days are used for the low-pass filtering, this method yields remarkably accurate results. 
{$b_{FIL}$ is calculated for each hemisphere of the 10 ensemble members  of CTL, and 95$\%$ confidence intervals are then calculated assuming these samples follow Gaussian distribution. }
The pressure-latitude pattern of eddy flux anomaly associated with the annular mode is also constructed by regressing low-pass filtered eddy fluxes onto the low-pass filtered zonal index, and the results compares well with those from LRF, with a pattern correlation exceeding 0.9. 

\begin{figure*}[t]
 \centerline{\includegraphics[width=0.5\textwidth]{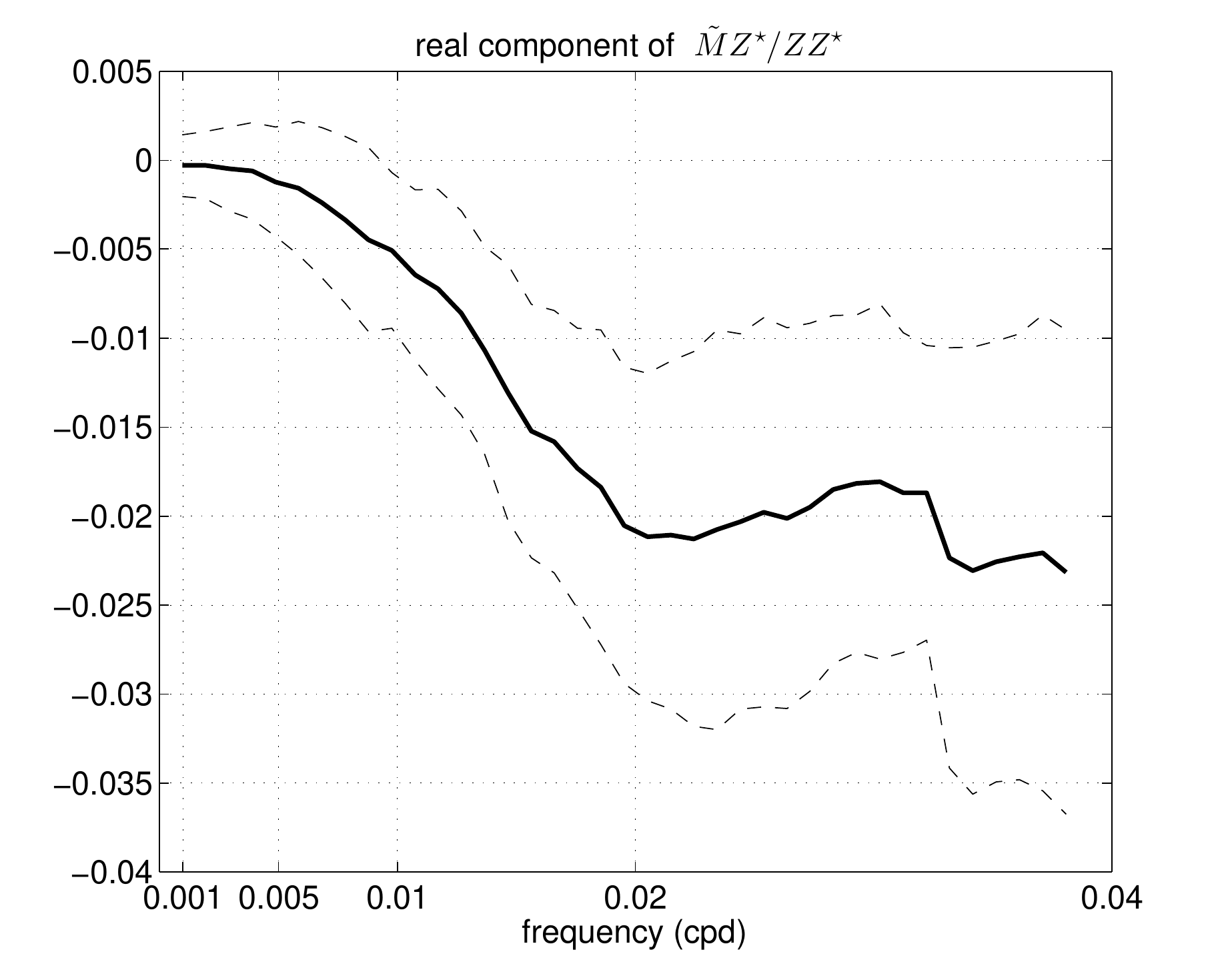}}
   \caption{Real component of $\frac{\tilde{M}Z^{*}} {ZZ^{*}}$ in CTL, with the dashed curves denoting 95$\%$ confidence intervals.}
  \label{fig12}
\end{figure*}

\subsection{Application to the reanalysis data}
The above three statistical methods are applied to estimate the strength of eddy-jet feedback in the reanalysis data, and the results are summarized in Figure 13.

\begin{figure*}[t]
 \centerline{\includegraphics[width=\textwidth]{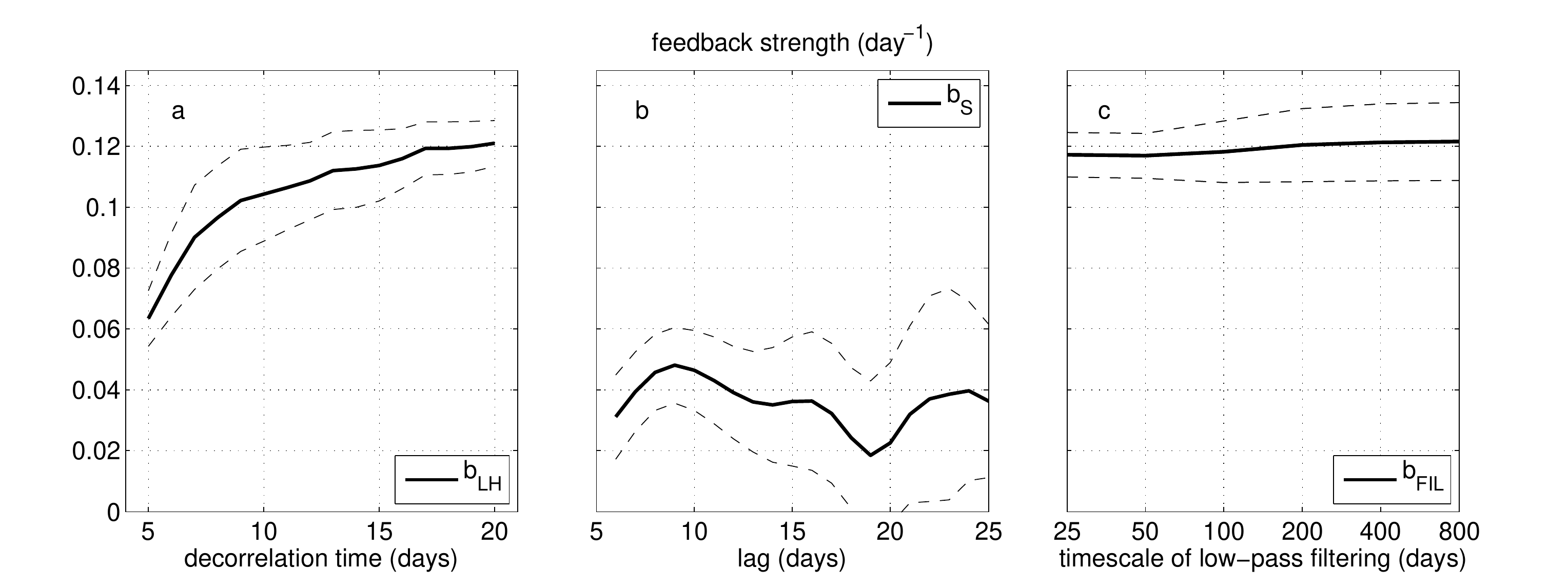}}
  \caption{Similar to Figure 8, except for the reanalysis data.}
  \label{fig13}
\end{figure*}

By minimizing the mean squared cross-correlations at lags longer than certain number of days as illustrated in Figure 4, $b_{LH}$ spans a range of values from around $0.06$ to $0.12$ day$^{-1}$ with the choices of decorrelation timescales of 5-20 days. 
The estimation for the reanalysis data is more sensitive to the choices of decorrelation and has larger margin of error compared to that of the idealized GCM (Figure 13a), which may partly be attributed to the shorter temporal length of the reanalysis data. 
Using lag regression, the estimated feedback strength is a function of lag days, and the margin of error grows with increasing lag (Figure 13b). Also, $b_{S}$ is more sensitive to the choices of lag days and has larger uncertainties than its counterpart with model outputs.

Although there is no ``ground truth" for the reanalysis data, the result obtained from regression with low-pass filtered data seems encouraging (Figure 13c). $b_{FIL}$ converges to around 0.121 day$^{-1}$ at low-frequency limit, which matches well with $b_{LH}$ with the decorrelation time of around 2 weeks. 
{There is also a significant contribution of medium-scales waves to total eddy forcing at intreaseasonal to interannual timescales in the reanalysis data (Figure 10b), and with daily data, $b_{FIL}$ is only around 0.053 day$^{-1}$.} 
The pattern of anomalous eddy fluxes associated with the annular mode is also calculated by regressing low-pass filtered time series (Figure 14). 
As expected, anomalous eddy flux converges zonal momentum into 60$^\circ$S-70$^\circ$S in the upper troposphere, and reinforces the anomalous zonal wind. 
Eddy anomalies originate from 60$^\circ$S-75$^\circ$S near the surface, where eddy heat flux is strengthened due to increased baroclinicity.

\begin{figure*}[t]
 \centerline{\includegraphics[width=\textwidth]{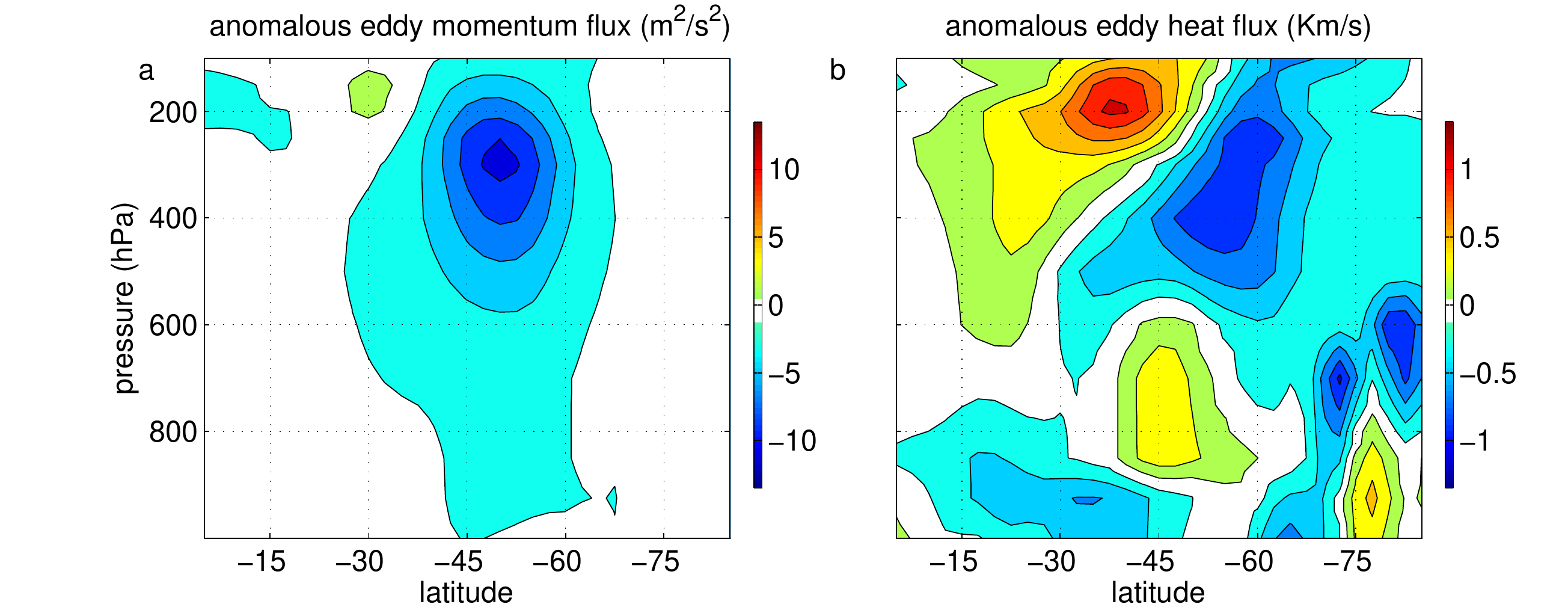}}
  \caption{Anomalous zonal average (a) eddy momentum flux and (b) eddy heat flux associated with the Southern annular mode in the reanalysis data.}
  \label{fig14}
\end{figure*}

While we do not have the LRF to separate out the mean-state-independent eddy forcing in the reanalysis, the low-pass filtering method only assumes that the mean-state-independent eddy forcing is sufficiently weak at the low-frequency limit so that the first term on the right hand side of Equation 9 is substantially smaller than the feedback factor $b$. 
Given that eddies are mostly generated at synoptic timescales, this seems a rather reasonable assumption. 
{A caveat of this assumption is that in the presence of an external low-frequency forcing (for example, due to stratospheric variability), the mean-state-independent eddy forcing might not be small at low frequencies (see an illustrative example in} \citet{Byrne2016} {and more discussions in the next section).}

\section{Discussions and summary}

The temporal persistence of the atmospheric annular mode has long been attributed to a positive eddy-jet feedback (e.g., Feldstein and Lee 1998; Robinson 2000; LH01), and statistical methods have been used to quantify the strength of the eddy feedback (LH01; S13). 
However, a recent study argues that one cannot discern the difference between the presence of an internal eddy feedback and external interrannual forcing using only the statistical methods \citep{Byrne2016}. 
Due to the stochastic nature of eddies, it is indeed impossible to separate the mean-state-dependent eddy flux from the mean-state-independent eddy flux and infer causality in the reanalysis data. 
In the present study, an LRF is used to identify the eddy response to anomalous mean flow associated with the annular mode in an idealized GCM, in which a positive eddy-jet feedback is confirmed unequivocally. 
With little spread across ten 44500-day integrations, EXP yields an eddy feedback strength of around 0.137 day$^{-1}$. 
{When the LRF is applied to the second EOF of zonal mean zonal wind, it yields a negative eddy feedback of -0.264 day$^{-1}$, consistent with the findings of LH01 who inferred the existence of a negative feedback in the second EOF of the observed Southern annular mode and attributed it to the barotropic governor effect (James 1987)}.

Using the LRF, the present study is able to provide a reasonably accurate estimation of the mean-state-independent eddy forcing. 
It is found that the spectral peak at synoptic timescales in the power spectrum of total eddy forcing ($m$) is dominated by the mean-state-independent eddy forcing ($\tilde{m}$). 
At intraseasonal and longer timescales, the amplitude of the mean-state-independent eddy forcing decreases with decreasing frequency{, and the total eddy forcing is dominated by mean-state-dependent eddy forcing. 
The role of the medium-scale waves on the annular mode is emphasized in the present study, which shows that less than half of the total eddy forcing can be captured using daily wind anomalies at interannual timescales as reported before in Kuroda and Mukougawa (2011). 
Without accounting for the medium-scales waves, the eddy feedback strength is underestimated by around 40$\%$.}

The present study focuses on an equinoctial mean state in the idealized GCM. While a number of previous studies \citep[e.g.,][]{Barnes2010, Byrne2016, Sheshadri2016} have brought attention to the seasonality of the annular mode. Seasonal variations of the persistence of the annular mode and eddy-jet feedback will be explored using the present methodology in a future study. 

The statistical methods proposed by LH01 and S13 are evaluated against the result from the LRF. 
By fitting the cross-correlations between the zonal index and eddy forcing as in LH01, the estimated feedback strength is fairly close to the result from the LRF, but it is difficult to determine \textit{a priori} an optimal value of decorrelation timescales, a parameter needed to calculate the best-fit.
Following S13, the output from lag-regression varies with lag days, and the feedback strength is largely underestimated, which suggests that the estimator is biased, and the assumption of S13 that the zonal index is decorrelated with the mean-state-independent eddy forcing beyond a lag time of a few days is not valid.
To be specific, the correlation between $\tilde{m}$ and $z$ cannot be neglected with a lag time spanning from a few days to as long as 30 days, as the mean-state-independent eddy forcing is quasi-oscillatory, with a broad peak in the power spectrum at synoptic timescale. 

To reduce the interference from the spectral peak of eddy forcing at synoptic timescales, we applied regressions on low-pass filtered eddy forcing and zonal index. 
The results converge to the value produced by the LRF when timescales longer than 200 days are used for the low-pass filtering. 
Given that the left hand side of Equation 4 is negligible at the low frequency limit, the fact that the power of the mean-state-independent eddy forcing is low at low frequencies implies that $b$ and $1/\tau$ are close to each other. 
The difference between $1/\tau$ and $b$, denoted as $1/\tau_{e}$, is constrained by examining $|Z / \tilde{M}|$, which can be derived from Equation 4:

\begin{equation}
\renewcommand{\theequation}{10}
\big|\frac{Z}{\tilde{M}} \big|= \big|\frac{1}{i\omega - 1/\tau_{e}}\big| = \frac{1}{\sqrt {\omega^2 + 1/\tau_{e}^2}}
\end{equation}

\noindent Taking advantage of the length of CTL, spectral analysis is conducted at very fine spectral resolution, i.e., 1/10000 cpd as in Figure 15.
At intraseasonal and shorter timescales, when $1/\tau_e$ is small compared to $\omega$, $|Z / \tilde{M}|$ is close to the $1/\omega$ curve (Figure 15). At the lowest frequencies, $|Z / \tilde{M}|$ is limited by $\tau_e$. 
The best-fit value of $\tau_e$ is 91 days from least squares fitting. The difference between $1/\tau$ and $b$ is smaller than 0.011 day$^{-1}$. 
The result is robust as $1/\tau_e$ ranges from 0.009 to 0.014 day$^{-1}$ when we applied least squares fitting to the ten ensemble members of CTL.
It leaves an intriguing question as to what physical processes determine the difference between $1/\tau$ and $b$, as $1/\tau$ and $b$ are connected, for example, via surface friction \citep{Chen2009}.

\begin{figure*}[t]
 \centerline{\includegraphics[width=25pc]{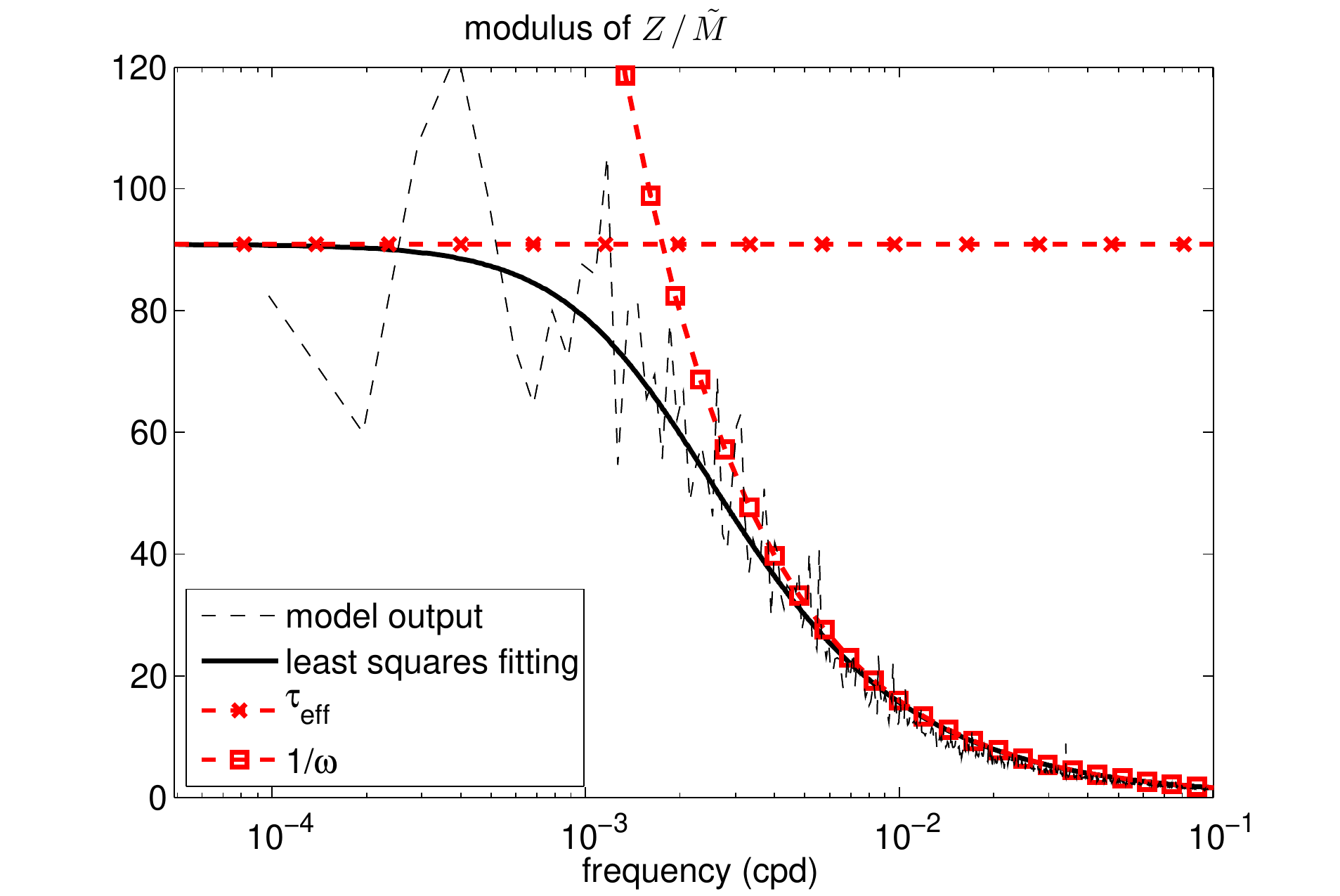}}
 \caption{Modulus of ${Z} /{\tilde{M}}$ from model outputs (black dashed curve) and least squares fitting (black solid curve) for model outputs of CTL.}
  \label{fig15}
\end{figure*}

When the statistical methods are applied to the reanalysis data, the performance of the methods proposed by LH01 and S13 is {influenced by the mean-state-independent eddy forcing}. For the reanalysis data, $b_{LH}$ and $b_{S}$ are more sensitive to the choices of parameters compared to their counterparts with model results. When the synoptic spectral peak is filtered out by low-pass filtering, with timescales longer than 200 days used for the low-pass filtering, $b_{FIL}$ converges to around 0.121 day$^{-1}$, which is close to the strength of eddy feedback in the idealized GCM. 

Although we cannot deny the presence of external eddy forcing at interannual timescale in the reanalysis data and its contribution to the persistence of the annular mode as suggested by \citet{Byrne2016}, the present study confirms the importance of a positive eddy-jet feedback to the persistence of the annular mode in an idealized GCM. 
The annular mode in this GCM compares well with that in reanalysis data, in terms of the spatial pattern of the leading EOF and the statistics of the zonal index and eddy forcing. 
The resemblance between the simulated annular mode and that in the reanalysis data suggests that the dry dynamical core with Held-Suarez physics, despite its idealized nature, is able to capture the essential dynamics of the annular mode. 
However, it should also be highlighted that the idealized model indeed produces a too persistent annular mode compared to the reanalysis. 
The eddy feedback is too strong in the idealized GCM, and it can be inferred that the difference between $1/\tau$ and $b$ is too small in the model. 
To what extent the results of the idealized GCM connect to the real atmosphere requires further research using observational data and a hierarchy of models. 

In addition, the present article provides another application of the LRF \citep{Hassanzadeh2015, Pedram2016a, Pedram2016b}. 
To quantify the strength of the eddy-jet feedback, one must be able to separate the anomalous eddies in response to a mean flow anomaly from the anomalous eddies that leads to the mean flow anomaly, which is difficult to do with statistical methods alone. 
Here the LRF is used to untangle the causal relationship in this eddy-jet feedback system, and provides the ``ground truth'' in the idealized GCM. Statistical methods are evaluated using model outputs, and then applied to the reanalysis data. 
The LRF can be calculated for GCMs of varying complexities, and the paradigm can be applied to a variety of problems involving identification of internal feedbacks.

\acknowledgments
This work is supported by NSF grants AGS-1062016 and AGS-1552385. The simulations are conducted on Harvard Odyssey cluster. 
Sincere thanks go to Nicholas Byrne for sharing analysis scripts and pointing out an important issue in an earlier draft of the manuscript that led us to realize the significance of using 6-hourly data. 
The authors thank Nicholas Byrne, Dennis Hartmann and Aditi Sheshadri for very constructive reviews and Martin Singh for discussions and comments on the manuscript.


 \bibliographystyle{ametsoc2014}
 \bibliography{eddy_clean}


\end{document}